\documentclass[conference]{IEEEtran}
\IEEEoverridecommandlockouts
\usepackage{cite}
\usepackage{amsmath,amssymb,amsfonts}
\usepackage{algorithmic}
\usepackage{textcomp}
\def\BibTeX{{\rm B\kern-.05em{\sc i\kern-.025em b}\kern-.08em
    T\kern-.1667em\lower.7ex\hbox{E}\kern-.125emX}}

\usepackage[nolist]{acronym}
\usepackage{multirow}
\usepackage{centernot}
\usepackage{steinmetz}
\usepackage{makecell}

\ifCLASSINFOpdf
\usepackage[pdftex]{graphicx}
\else
\fi

\ifCLASSOPTIONcompsoc
\usepackage[caption=false,font=normalsize,labelfont=sf,textfont=sf]{subfig}
\else
\usepackage[caption=false,font=footnotesize]{subfig}
\fi

\usepackage[dvipsnames]{xcolor}
  
\def\ans{\textcolor{Black}} %answering the reviewer

\usepackage{bm}
\newcommand{\mtx}[1]{\bm{\mathbf{#1}}}

\acused{AC}
\acused{LP-DOE}
\acused{LP-RDOE}
\acused{NLP-DOE}
\acused{NLP-RDOE}

\begin{acronym}
	\acro{13NF}{IEEE 13 Node Test Feeder}
	\acro{123NF}{IEEE 123 Node Test Feeder}
    \acro{AEE}{active energy entities}
	\acrodefplural{AEE}[AEEs]{active energy entities}
	\acro{AC}{alternating current}
	\acro{CIM}{current injection method}
	\acro{DER}{distributed energy resource}
	\acrodefplural{DER}[DERs]{distributed energy resources}
    \acro{DOE}{dynamic operating envelope}
	\acrodefplural{DOE}[DOEs]{dynamic operating envelopes}
    \acro{DSO}{distribution system operator}
	\acrodefplural{DSO}[DSOs]{distribution system operators}
    \acro{EV}{electric vehicle}
	\acrodefplural{EV}[EVs]{electric vehicles}
	\acro{ESS}{energy storage system}
	\acrodefplural{ESS}[ESSs]{energy storage systems}
	\acro{IEC}{International Electrotechnical Commission}
    \acro{LACE}{Linear Analytical Calculation of Envelopes}
    \acro{LP}{linear programming}
    \acro{LP-DOE}{LP-DOE}
    \acro{LP-RDOE}{LP-RDOE}
	\acro{L-QP}{Local-Quadratic Program}
	\acro{LV}{low-voltage}
	\acro{MV}{medium-voltage}
    \acro{MPC}{model predictive control}
    \acro{NLP}{non-linear programming}
    \acro{NLP-DOE}{NLP-DOE}
    \acro{NLP-RDOE}{NLP-RDOE}
	\acro{OPF}{optimal power flow}
	\acro{OPF-PC}{OPF-based Proportional Control}
	\acro{PCC}{point of common coupling}
	\acro{PF}{power flow}
	\acro{PV}{photovoltaic}
	\acro{QP}{quadratic program}
	\acro{QCQP}{quadratically constrained quadratic program}
    \acro{RHO}{receding-horizon optimization}
	\acro{SOC}{state of charge}
	\acro{TOU}{time-of-use}
	\acro{VB-VVW}{Voltage-Based Volt-Var-Watt}
\end{acronym}

\begin{document}

\title{Robust Dynamic Operating Envelopes in \\ Unbalanced Three-Phase Distribution Systems
%\vspace{-0.5em}
\thanks{© 2026 IEEE.  Personal use of this material is permitted.  Permission from IEEE must be obtained for all other uses, in any current or future media, including reprinting/republishing this material for advertising or promotional purposes, creating new collective works, for resale or redistribution to servers or lists, or reuse of any copyrighted component of this work in other works.}
}

\author{
\makebox[\linewidth][c]{%
  \begin{tabular}{cc}
    \begin{minipage}{0.47\linewidth}\centering
      Wilhiam~de~Carvalho \\
      Luxembourg Institute of Science and Technology\\
      Esch-sur-Alzette, Luxembourg
    \end{minipage}
    &
    \begin{minipage}{0.47\linewidth}\centering
      Florin~Capitanescu\\
      Luxembourg Institute of Science and Technology\\
      Esch-sur-Alzette, Luxembourg
    \end{minipage}
  \end{tabular}
} \\ \\
\makebox[\linewidth][c]{%
  \begin{tabular}{ccc}
    \begin{minipage}{0.3\linewidth}\centering
      Cyril~Rasic\\
      The University of Mons\\
      Mons, Belgium
    \end{minipage}
    &
    \begin{minipage}{0.35\linewidth}\centering
      Jean-François~Toubeau\\
      The University of Mons\\
      Mons, Belgium
    \end{minipage}
    &
    \begin{minipage}{0.3\linewidth}\centering
      François~Vallée\\
      The University of Mons\\
      Mons, Belgium
    \end{minipage}
  \end{tabular}
}
}

\IEEEaftertitletext{\vspace{-0pt}}

\maketitle

\begin{abstract}
This paper proposes a robust optimization formulation to calculate dynamic operating envelopes (DOEs) to safely operate unbalanced three-phase distribution systems. Unlike conventional formulations that satisfy network constraints only at the envelope bound, the robust formulation covers the entire envelope range. We formulate a robust non-linear programming (NLP) problem with the full AC power flow equations, as well as an approximate linear programming (LP) model. Numerical simulations are run with real-world data from Belgium and two different distribution test feeders. The paper compares the conventional approaches with their robust counterparts and examines the trade-off between constraint violation and envelope size as well as accuracy and solve time aspects.
\end{abstract}


\section{Introduction}

%1st: Intro: motivate DOEs - DER challenge; DOE is key
With the drastic uptake of \acp{DER} in distribution systems, there is a greater need for coordinating such assets to safely maximize network utilization. The recently proposed \ac{DOE} \cite{petrou2021} has emerged as a promising tool to account for volatile and \ac{DER}-rich distribution grids. \ac{DOE} calculation engines find the available grid capacity at a given time to allocate in the form of envelopes to \acp{AEE} across the network~\cite{unimelb2024arthur}.

%2nd: DOE concept and monotonicity
Conventional \ac{DOE} engines compute two key optimal points to form the envelope: the maximum load import (demand) and maximum export (generation). The underlying assumption is that the network constraints will then be feasible provided every \ac{AEE} operates within those two given bounds. Fundamentally, this translates into assuming a monotonic \ans{property} such that larger import from any \ac{AEE} always results in larger voltage drops/line power flows --- and vice versa \cite{moring2024}.

%3rd: Motivating our paper: DOE problem - works for 1ph, but not 3ph
Monotonicity has shown to hold empirically with single-phase (balanced) models, and it can be proven with balanced linear models. However, distribution networks present significant coupling between phases and unbalanced levels, requiring accurate three-phase models to capture the physics of the grid. The coupling between phases leads to non-monotonic behavior and hence conventional \ac{DOE} approaches do not necessarily satisfy grid operation constraints (e.g. voltage, thermal), even with \acp{AEE} operating well within their \ac{DOE} bounds \cite{moring2024, binliu2022sensi}.

%4th: literature review and gap
Recent works have identified this phenomenon and proposed \ac{DOE} approaches \ans{based on linear power flow models} to circumvent such challenges \cite{binli2024robust, russell2023robust}. \ans{The robust approach in \cite{binli2024robust} requires three distinct stages, with one stage still resulting in a} non-convex non-linear optimization problem, lacking both accuracy and scalability. A robust \ac{LP} approach is provided in \cite{russell2023robust} to address issues with voltage phase unbalancing and magnitude, disregarding accurate non-linear models and thermal capacity of distribution grids. Other papers \cite{michliu2022, petrou2021, petrou2020isgt, unimelb2023fairnessrep, michliu2021market} have used three-phase models in their \ac{DOE} optimization, without addressing the non-monotonic \ans{relationship} arising from phase coupling.

%5th: our paper: contributions?
The original contribution of this paper is robust optimization formulations\ans{, with both linear and non-linear power flow models,} to calculate \acp{DOE} in unbalanced three-phase distribution systems. The robust formulation exploits the root cause of non-monotonicity and considers uncertainty in the realized power of \acp{AEE} to satisfy voltage and thermal constraints for all scenarios. Our numerical simulations examine trade-offs between robust and conventional \acp{DOE}, as well as linear and non-linear models \ans{for \ac{DOE} calculation}.

\section{Preliminaries}

\begin{figure}[!t]
	\centering
	\includegraphics[width=0.90\linewidth]{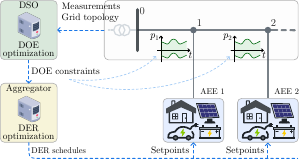}
	\caption{\ans{\ac{DOE} Overview. The \ac{DSO} keeps the network secure by computing and providing dynamic bounds (i.e., \acp{DOE}) to aggregators or the prosumers themselves (i.e., \acp{AEE}). Flexible \acp{DER} are then optimized by aggregators or \acp{AEE} to their desired objectives, subject to the given \acp{DOE}.}}
	\vspace{-0pt}
	\label{fig:DOE_overview}
\end{figure}

As in Fig.~\ref{fig:DOE_overview}, let node~0 denote the substation (slack) bus, $\mathcal{N}:=\{1,2,...,N\}$ the set of downstream nodes and $\mathcal{E}$ the set of all line segments of an unbalanced three-phase distribution network. For every branch $(m,n) \in \mathcal{E}$ and phase $\phi \in \{a,b,c\}$, denote by $I_{mn}^\phi$ the complex current, $P_{mn}^\phi$ the real power, and $Q_{mn}^\phi$ the reactive power. For every node $m$ and phase $\phi \in \{a,b,c\}$, denote by $V_{m}^\phi$ the complex voltage, and $\widetilde{p}_{m}^\phi$ and $\widetilde{q}_{m}^\phi$ the known non-controllable real and reactive load, respectively. The decision variable of interest (envelope bound), is represented by the controllable real power component, denoted by $p_{m}^\phi$.

We collect the scalars in vectors as for example in $\mtx{V} = [V_0^a, V_0^b, V_0^c, V_1^a, ..., V_N^c]^\top$, and $\mtx{V}_{re}$ and $\mtx{V}_{im}$ collects the real and imaginary part, respectively, of the concerning matrix/vector. The three-phase system-wide Y-bus matrix is denoted by its real ($\mtx{G}$) and imaginary ($\mtx{B}$) part. The full AC power flow equations based on the current-injection (I-V) model \cite{molzahn2019mono} are formulated in rectangular coordinate as:
\begin{equation}
\mtx{I}_{re} + \mtx{G} \mtx{V}_{re} - \mtx{B} \mtx{V}_{im} = \mtx{0},
\label{eq:pf_bim_ireal}
\end{equation}
\begin{equation}
\mtx{I}_{im} + \mtx{B} \mtx{V}_{re} + \mtx{G} \mtx{V}_{im} = \mtx{0},
\label{eq:pf_bim_iimag}
\end{equation}
\begin{equation}
\mtx{p} + \widetilde{\mtx{p}} - \mtx{V}_{re} \mtx{I}_{re} - \mtx{V}_{im} \mtx{I}_{im} = \mtx{0},
\label{eq:pf_bim_pnode}
\end{equation}
\begin{equation}
\widetilde{\mtx{q}} + \mtx{V}_{re} \mtx{I}_{im} - \mtx{V}_{im} \mtx{I}_{re} = \mtx{0}.
\label{eq:pf_bim_qnode}
\end{equation}

Optimization engines with the power flow equations \eqref{eq:pf_bim_ireal}-\eqref{eq:pf_bim_qnode} results in non-convex \ac{NLP}. This often motivates the formulation of linear equations to support convex \ac{DOE} optimization problems \cite{michliu2022}. We adopt the approximate linear equations based on the three-phase branch flow model derived in \cite{sankur2019phasor}, given by
\begin{equation}
\mtx{P}_{lm} = \mtx{p}_{m} + \widetilde{\mtx{p}}_{m} + \sum\nolimits_{\scriptscriptstyle n:(m,n) \in \mathcal{E}} \mtx{P}_{mn}, \ \ \ \forall (l,m) \in \mathcal{E},
\label{eq:bfm_p}
\end{equation}
\begin{equation}
\mtx{Q}_{lm} = \widetilde{\mtx{q}}_{m} + \sum\nolimits_{\scriptscriptstyle n:(m,n) \in \mathcal{E}} \mtx{Q}_{mn},  \ \ \ \ \ \ \ \ \ \ \forall (l,m) \in \mathcal{E},
\label{eq:bfm_q}
\end{equation}
\begin{equation}
\mtx{U}_{m} = \mtx{U}_{n} + \mtx{M}_{mn} \mtx{P}_{mn} + \mtx{N}_{mn} \mtx{Q}_{mn},  \ \ \ \forall (m,n) \in \mathcal{E},
\label{eq:bfm_v}
\end{equation}
where $U_m^\phi = |V_m^\phi|^2$ is the voltage magnitude square,
\begin{equation}
\mtx{M}_{mn} = \left[ \begin{smallmatrix}
2r_{mn}^{aa}                        & -r_{mn}^{ab}+\sqrt{3} x_{mn}^{ab}     & -r_{mn}^{ac}-\sqrt{3} x_{mn}^{ac}   \\
-r_{mn}^{ba}-\sqrt{3} x_{mn}^{ba}   & 2r_{mn}^{bb}                          & -r_{mn}^{bc}+\sqrt{3} x_{mn}^{bc}   \\
-r_{mn}^{ca}+\sqrt{3} x_{mn}^{ca}   & -r_{mn}^{cb}-\sqrt{3} x_{mn}^{cb}     & 2r_{mn}^{cc}
\end{smallmatrix} \right] ,
\label{eq:bfm_m}
\end{equation}
\begin{equation}
\mtx{N}_{mn} = \left[ \begin{smallmatrix}
2x_{mn}^{aa}                        & -x_{mn}^{ab}-\sqrt{3} r_{mn}^{ab}     & -x_{mn}^{ac}+\sqrt{3} r_{mn}^{ac}   \\
-x_{mn}^{ba}+\sqrt{3} r_{mn}^{ba}   & 2x_{mn}^{bb}                          & -x_{mn}^{bc}-\sqrt{3} r_{mn}^{bc}   \\
-x_{mn}^{ca}-\sqrt{3} r_{mn}^{ca}   & -x_{mn}^{cb}+\sqrt{3} r_{mn}^{cb}     & 2x_{mn}^{cc}
\end{smallmatrix} \right] ,
\label{eq:bfm_n}
\end{equation}
and $r_{mn}^{\phi \psi}$ is the resistance and $x_{mn}^{\phi \psi}$ the reactance of the three-phase line segment $(m,n)$. Matrices $\mtx{M}_{mn}, \mtx{N}_{mn}$ represent the coefficients in which power impacts the voltage deviation, with the diagonal entries reflecting series line impedance, and off diagonal the mutual impedances. We consider these coefficients static as in \eqref{eq:bfm_m}, \eqref{eq:bfm_n}, noting that $\mtx{M}_{mn}, \mtx{N}_{mn}$ can be regularly updated with sequential linearization techniques. 

Operating envelopes are obtained by computing the additional amount of power $\mtx{p}$ that nodes can absorb (import case) or inject (export case) before breaching the network voltage and thermal constraints. That is, \ac{DOE} optimization engines compute the spare grid capacity that can be provided to individual \acp{AEE} in the form of power envelopes (see Fig.~\ref{fig:DOE_overview}). \ans{Similar to \cite{moring2024, michliu2022, petrou2020isgt}}, the conventional optimization problem to calculate the maximum size of import \acp{DOE} is written as
\begin{subequations}
\begin{align}
\max_{\substack{\mtx{p}}} \quad  & \mtx{1}^\top \mtx{p} \\
\textrm{s.t.}
\quad & \eqref{eq:pf_bim_ireal}-\eqref{eq:pf_bim_qnode} \mbox{ or }  \eqref{eq:bfm_p}-\eqref{eq:bfm_v}, \\
\quad & \mtx{P}_{01}^2 + \mtx{Q}_{01}^2 \leq \overline{\mtx{S}}_{01}^2, \\
\quad & \underline{\mtx{U}} \leq \mtx{U} \leq \overline{\mtx{U}}, \\
\quad & \underline{\mtx{p}} \leq \mtx{p} \leq \overline{\mtx{p}} , \\
\quad & \mtx{p} \geq \mtx{0} ,
\end{align}
\label{eq:conv_lp-nlp-doe}
\end{subequations}
\hspace{-7pt} where the square and inequality in (\ref{eq:conv_lp-nlp-doe}c) are applied element-wise. The voltage at the slack node is fixed at a known reference/measured value, adding an equality constraint for $\mtx{V}_{0,re}, \mtx{V}_{0,im}$ when using \eqref{eq:pf_bim_ireal}-\eqref{eq:pf_bim_qnode} or for $\mtx{U}_0$ when using \eqref{eq:bfm_p}-\eqref{eq:bfm_v}. The objective function in \eqref{eq:conv_lp-nlp-doe} maximizes the combined additional power that nodes than consume (import). Constraint (\ref{eq:conv_lp-nlp-doe}b) represent the adopted power flow model, (\ref{eq:conv_lp-nlp-doe}c) is the thermal limit at the feeder head, (\ref{eq:conv_lp-nlp-doe}d) the voltage limits, (\ref{eq:conv_lp-nlp-doe}e) the physical limits of participant \acp{DER}, and (\ref{eq:conv_lp-nlp-doe}f) reflects the import case. The export \ac{DOE} problem is formulated by simply changing the objective to $\min$ and (\ref{eq:conv_lp-nlp-doe}f) to $\mtx{p}\leq\mtx{0}$.

With the linear power flow model, (\ref{eq:conv_lp-nlp-doe}c) turns into a linear inequality and optimization \eqref{eq:conv_lp-nlp-doe} boils down to a \ac{LP} problem. We name the conventional optimization \eqref{eq:conv_lp-nlp-doe} as \ac{LP-DOE} when adopting the linear model \eqref{eq:bfm_p}-\eqref{eq:bfm_v}, and \ac{NLP-DOE} with the non-linear model \eqref{eq:pf_bim_ireal}-\eqref{eq:pf_bim_qnode} -- both used as benchmark approaches in this paper. By solving problem \eqref{eq:conv_lp-nlp-doe}, the \ac{DSO} or an equivalent service provider sends the optimal bounds $(p_m^\phi)^*$ to individual \acp{AEE} across the network (or their respective aggregators). \acp{AEE} then are allowed to optimize their local \acp{DER} however they like as long as they keep their import/export power within the given bounds.

The conventional \ac{DOE} optimization problem maximizes the upper bound of the envelope assuming that any value between $\mtx{0}$ and the optimal upper bound ($\mtx{p}^*$) results in feasible network operation. This fundamental monotonicity assumption can be proven in single-phase (balanced) linear power flow models and empirically holds for non-linear single-phase models \cite{moring2024}. However, the coupling effect of mutual impedances in three-phase distribution grids and highly unbalanced loading conditions explicitly breaks the monotonicity relationship between nodal power and voltage and thermal constraints \cite{binliu2022sensi}.

It becomes clear from $\mtx{M}_{mn}$ in \eqref{eq:bfm_v} that positive and negative coefficients make the real power of different phases to impact voltage magnitude in opposite directions (drop/rise) \cite{binliu2022sensi}. The same phenomenon is also observed with the non-linear three-phase model for \ac{MV} and \ac{LV} networks and for both overhead and underground lines. Without monotonicity, simply maximizing the upper bound does not guarantee the entire \ac{DOE} range $\mtx{0} \leq \mtx{p} \leq \mtx{p}^*$ is feasible.

\subsection{Conceptual and Geometrical Understanding}

\begin{figure}[!t]
	\centering
	\subfloat[]{%
		\includegraphics[width=0.33\linewidth]{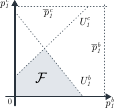}}
	\subfloat[]{%
		\includegraphics[width=0.33\linewidth]{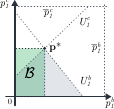}}
    \subfloat[]{%
		\includegraphics[width=0.33\linewidth]{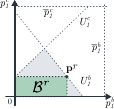}}
	\caption{Import \ac{DOE} calculation with the linear model: (a) network feasible region; (b) conventional \ac{DOE} optimization; and (c) robust \ac{DOE} optimization.}
	\label{fig:feasible_reg}
\end{figure}

We provide here a geometrical interpretation of the non-monotonic behavior similar to \cite{binliu2022sensi}. Fig.~\ref{fig:feasible_reg} illustrates a \ac{DOE} calculation case with the \ac{LP-DOE} engine for a three-phase two-node feeder, with phase~b and c participating in the \ac{DOE} program. Fig.~\ref{fig:feasible_reg}(a) depicts the network feasible region $\mathcal{F}$ where $p_1^b$ and $p_1^c$ are allowed to vary without breaching network constraints. Due to the negative coupling ($M_{01}^{cb}$), $\mathcal{F}$ is characterized as not-downward closed in the $p_1^b$ direction, such that not every box (i.e., envelope) closed at the edge of $\mathcal{F}$ is entirely inscribed in it.

Fig.~\ref{fig:feasible_reg}(b) illustrates the optimal solution ($\mtx{p}^*$) of problem \eqref{eq:conv_lp-nlp-doe}, and the import envelopes represented by the box region $\mathcal{B}$. Observe that the voltage magnitude c does not comply with the constraint for all power values within the envelope $\mathcal{B}$. For any realization of $p_1^b < (p_1^b)^*$, the actual feasible upper bound of $p_1^c$ is significantly smaller than $(p_1^c)^*$.

Fig.~\ref{fig:feasible_reg}(c) depicts a robust \ac{DOE} calculation that maximizes the combined upper bound of envelopes, provided the box is fully inscribed in $\mathcal{F}$. In this sense, we propose a robust \ac{DOE} approach that accounts for uncertainties in power realization, such that any realized power within the envelope $\mtx{0} \leq \mtx{p} \leq \mtx{p}^r$ does not breach thermal or voltage limits, maintaining a secure and efficient network operation.

%Our proposed Robust DOE approach
\section{Problem Formulation}

We start by reformulating the linear \ac{DOE} to understand and address the non-monotonic constraints. Then we translate the mathematical insights developed in the linear problem to the \ac{NLP} case, and provide a practical conservative approach to obtain robust \ac{DOE} values with the exact non-linear model.

Similar to \cite{decarvalho2024arxiv, decarvalho2026cired}, we define the voltage magnitude deviation w.r.t node~0 as $E_m^\phi := U_0^\phi - U_m^\phi$, and collect all scalars in the vector $\mtx{E}$. Define system-wide matrices $\mtx{R} := \mtx{A} \mbox{diag}(\mtx{M) A^\top}$ and $\mtx{X} := \mtx{A} \mbox{diag}(\mtx{N) A^\top}$, where $\mtx{A}$ is the three-phase path incidence matrix, and $\mtx{M}$, $\mtx{N}$ stack the $3\times3$ blocks $\mtx{M}_{mn}$, $\mtx{N}_{mn}$, respectively, for all line segments. The resulting linear power flow model is given by
\begin{equation}
\mtx{E} = \mtx{R} \mtx{p} + \widetilde{\mtx{E}} ,
\label{eq:compact_lindist}
\end{equation}
where $\widetilde{\mtx{E}} = \mtx{R} \widetilde{\mtx{p}} + \mtx{X} \widetilde{\mtx{q}}$ represents the known underlying voltage deviation due to non-controllable (base) load. The linear power flow \eqref{eq:compact_lindist} is now w.r.t. nodal power and the coefficients of matrix $\mtx{R}$ directly dictates how the \ac{DOE} of each node on the network impacts the voltage. Matrix $\mtx{R}$ represents the path impedance from node~0 to \ans{every} other node in the network and, similar to \eqref{eq:bfm_m}, contains positive and negative coefficients of phase-coupled impedances.

Next, we define voltage constraints for \eqref{eq:compact_lindist}, \ans{adopting the positive voltage drop convention}. Denote by $\overline{\mtx{E}}:= \mtx{U}_0 - \underline{\mtx{U}}$ the largest voltage drop allowed and by $\underline{\mtx{E}}:=\mtx{U}_0 - \overline{\mtx{U}}$ the \ans{smallest voltage drop} allowed. Finally, we define the spare voltage drop by $\widehat{\mtx{E}}:=\overline{\mtx{E}}-\widetilde{\mtx{E}}$ and the spare voltage rise by $\breve{\mtx{E}}:=\underline{\mtx{E}}-\widetilde{\mtx{E}}$ to write the \ans{voltage} constraint \ans{for \eqref{eq:compact_lindist}} as
\begin{equation}
\breve{\mtx{E}} \leq \mtx{R} \mtx{p} \leq \widehat{\mtx{E}} .
\label{eq:compact_lindist_vconst}
\end{equation}
Similarly, to reformulate (\ref{eq:conv_lp-nlp-doe}c) explicitly in terms of $\mtx{p}$ as in \cite{decarvalho2026cired}, let the non-controllable real power at the feeder head be $\widetilde{P}_{01}^\phi = \sum_{\scriptscriptstyle m \in \mathcal{N}} \widetilde{p}_{m}^\phi$ and the reactive power be $\widetilde{Q}_{01}^\phi = \sum_{\scriptscriptstyle m \in \mathcal{N}} \widetilde{q}_{m}^\phi$. We obtain the real power limit at the feeder head by $\overline{P}_{01}^\phi = ((\overline{S}_{01}^\phi)^2 - (\widetilde{Q}_{01}^\phi)^2)^{1/2}$, where $\overline{S}_{01}^\phi$ is the nominal apparent power of the substation transformer. Finally, we denote the spare thermal limit by $\widehat{P}_{01}^\phi := \overline{P}_{01}^\phi - \widetilde{P}_{01}^\phi$ to write the thermal constraint for each phase as
\begin{equation}
\mtx{1}^\top \mtx{p}^{\phi} \leq \widehat{P}_{01}^{\phi} .
\label{eq:compact_lindist_pconst}
\end{equation}

\subsection{Robust Three-Phase \ac{DOE} - \ac{LP} Model}

We propose a robust optimization counterpart to compute the largest combined power, such that the envelope is fully inscribed in the feasible region. The robust formulation takes into account 100\% uncertainty in the realized power ($u$) within the import envelope $0 \leq u \leq p$. Mathematically, we write the import robust \ac{DOE} as
\begin{subequations}
\begin{align}
\max_{\substack{\mtx{u}, \mtx{p}}} \quad  & \mtx{1}^\top \mtx{p} \\
\textrm{s.t.}
\quad & \forall \mtx{u} \in \ \mathcal{U}(\mtx{p}) = \{\mtx{u} : \mtx{0} \leq \mtx{u} \leq \mtx{p}\} : \\
\quad & \breve{\mtx{E}} \leq \mtx{R} \mtx{u} \leq \widehat{\mtx{E}} , \\
\quad & \mtx{1}^\top \mtx{u}^{\phi} \leq \widehat{P}_{01}^{\phi}, \ \ \ \forall \phi \in \{a,b,c \}, \\
\quad & \underline{\mtx{p}} \leq \mtx{u} \leq \overline{\mtx{p}} .
\end{align}
\label{eq:robust_lp-doe}
\end{subequations}
\hspace{-5pt} where $\mathcal{U}(\mtx{p})$ is the uncertainty set that represents the envelope itself. Note that \eqref{eq:robust_lp-doe} maximizes the upper bound of the import envelope and the constraints are formulated to make sure the problem is feasible for all realized power within the envelope. Constraints (\ref{eq:robust_lp-doe}d), (\ref{eq:robust_lp-doe}e) are monotonic, since (element-wise) $\mtx{u}_{(1)} \leq \mtx{u}_{(2)} \implies f(\mtx{u}_{(1)}) \leq f(\mtx{u}_{(2)}), \ \forall \mtx{u}_{(1)}, \mtx{u}_{(2)} \in \mathcal{U}(\mtx{p})$. In this sense, we can equivalently write these constraints in terms of $p$ as any power realization within the import envelope will also satisfy these constraints.

In contrast, the relationship \ans{in (\ref{eq:robust_lp-doe}c)} between the voltage magnitude and import power is not monotonic as a power realization that is smaller than the upper envelope not necessarily guarantees feasibility (see Fig.~\ref{fig:feasible_reg}). Mathematically, it holds that $\mtx{u}_{(1)} \leq \mtx{u}_{(2)} \centernot\implies \mtx{R}\mtx{u}_{(1)} \leq \mtx{R}\mtx{u}_{(2)}$. \ans{Applying the worst-case operator as in \cite{russell2023robust, bental2021book}, constraint (\ref{eq:robust_lp-doe}c) is written~as}
\begin{align}
    \max_{\substack{\mtx{0}\leq \mtx{u} \leq \mtx{p}}} (\mtx{R} \mtx{u}) \leq \widehat{\mtx{E}}, \ \ \ \ \ \ \
    \min_{\substack{\mtx{0}\leq \mtx{u} \leq \mtx{p}}} (\mtx{R} \mtx{u}) \geq \breve{\mtx{E}} .
    \label{eq:rob_maxmin_const}
\end{align}
Note that in \eqref{eq:rob_maxmin_const} the maximum occurs when $u=p$ for all positive and $u=0$ for all negative coefficients of $\mtx{R}$. Alternatively, the minimum is achieved when $u=p$ for all negative and $u=0$ for all positive coefficients of $\mtx{R}$. Let $\mtx{R} = \mtx{R}^+ + \mtx{R}^-$, where entries $R_{mn}^+ = \max(R_{mn}, 0)$ and $R_{mn}^- = \min(R_{mn}, 0)$, $\forall m,n$ \cite{bental2021book, russell2023robust}. For completeness, we then reformulate the robust problem with the deterministic equivalent as
\begin{subequations}
\begin{align}
\max_{\substack{\mtx{p}}} \quad  & \mtx{1}^\top \mtx{p} \\
\textrm{s.t.}
\quad & \mtx{R}^+ \mtx{p} \leq \widehat{\mtx{E}} , \\
\quad & \mtx{R}^- \mtx{p} \geq \breve{\mtx{E}} , \\
\quad & \mtx{1}^\top \mtx{p}^{\phi} \leq \widehat{P}_{01}^{\phi}, \ \ \ \forall \phi \in \{a,b,c \},\\
\quad & \underline{\mtx{p}} \leq \mtx{p} \leq \overline{\mtx{p}} , \\
\quad & \mtx{p} \geq \mtx{0} .
\end{align}
\label{eq:robust_R+_lp-doe}
\end{subequations}
\hspace{-5pt} We term \eqref{eq:robust_R+_lp-doe} as the robust linear \ac{DOE} or simply \ac{LP-RDOE}. Note that \eqref{eq:robust_R+_lp-doe} is a single-level tractable \ac{LP} problem with many  available solvers, such as CPLEX, Gurobi, GLPK that are efficient and reliable for real-world applications. Similar robust optimization is formulated for the export \ac{DOE}.

\subsection{Robust Three-Phase \ac{DOE} - \ac{NLP} Model}

The robust \ac{DOE} requires that constraints are satisfied for all power realizations in the uncertainty set $\mathcal{U}(\mtx{p})$. With the linear model, the robust counterpart has a closed-form formulation that turns into a single \ac{LP} problem \eqref{eq:robust_R+_lp-doe}. With the non-linear equations, analytically finding the worst-case within the entire uncertainty set is not trivial. We have observed that phase coupling as in \eqref{eq:bfm_m} has a major impact on the non-monotonic behavior of the non-linear constraints. Considering phases have opposite contributions on the voltage magnitude, it is crucial to take into account different phase-wise realization scenarios when computing \acp{DOE}.

Based on this knowledge of the physics of three-phase networks, we select critical edge-cases to represent the worst cases of $\mathcal{U}(\mtx{p})$. We define the discrete uncertainty set by $\mathcal{D}(\mtx{p}) = \{(\mtx{p}^a, \mtx{p}^b, \mtx{p}^c), \allowbreak (\mtx{0}, \mtx{p}^b, \mtx{p}^c), \allowbreak (\mtx{p}^a, \mtx{0}, \mtx{p}^c), \allowbreak (\mtx{p}^a, \mtx{p}^b, \mtx{0}), \allowbreak (\mtx{p}^a, \mtx{0}, \mtx{0}), \allowbreak (\mtx{0}, \mtx{p}^b, \mtx{0}), \allowbreak (\mtx{0}, \mtx{0}, \mtx{p}^c) \}$. For completeness, we formulate the import robust \ac{NLP} counterpart as
\begin{subequations}
\begin{align}
\max_{\substack{\mtx{u}, \mtx{p}}} \quad  & \mtx{1}^\top \mtx{p} \\
\textrm{s.t.}
\quad & \forall \mtx{u} \in \ \mathcal{D}(\mtx{p}) : \\
\quad & \mtx{I}_{re} + \mtx{G} \mtx{V}_{re} - \mtx{B} \mtx{V}_{im} = \mtx{0}, \\
\quad & \mtx{I}_{im} + \mtx{B} \mtx{V}_{re} + \mtx{G} \mtx{V}_{im} = \mtx{0}, \\
\quad & \mtx{u} + \widetilde{\mtx{p}} - \mtx{V}_{re} \mtx{I}_{re} - \mtx{V}_{im} \mtx{I}_{im} = \mtx{0}, \\
\quad & \widetilde{\mtx{q}} + \mtx{V}_{re} \mtx{I}_{im} - \mtx{V}_{im} \mtx{I}_{re} = \mtx{0}, \\
\quad & \mtx{P}_{01}^2 + \mtx{Q}_{01}^2 \leq \overline{\mtx{S}}_{01}^2, \\
\quad & \underline{\mtx{U}} \leq \mtx{V}_{re}^2 + \mtx{V}_{im}^2 \leq \overline{\mtx{U}}, \\
\quad & \underline{\mtx{p}} \leq \mtx{u} \leq \overline{\mtx{p}} , \\
\quad & \mtx{p} \geq \mtx{0} .
\end{align}
\label{eq:robust_D_nlp-doe}
\end{subequations}
\hspace{-7pt} where the square and inequality in (\ref{eq:robust_D_nlp-doe}g), (\ref{eq:robust_D_nlp-doe}h) are applied element-wise. We solve this \ac{NLP} optimization by running it seven different times, i.e., one for each element in the set $\mathcal{D}(\mtx{p})$, and selecting the most conservative envelope values.

\ans{This physics-based heuristic approach has shown to provide grid-safe \acp{DOE} in all our numerical tests, although other edge-cases might still be possible in extreme cases due to the non-linear nature of the grid.} We term \eqref{eq:robust_D_nlp-doe} as the \ac{NLP-RDOE} approach. A similar robust problem is formulated for the export \ac{DOE}.

\section{Numerical Simulations}

We run numerical simulations with a conceptual and a realistic three-phase test feeder. Both import and export \acp{DOE} are computed and all 4 approaches are compared. The conventional \ac{LP-DOE} is obtained by solving \eqref{eq:conv_lp-nlp-doe} with the linear equations \eqref{eq:bfm_p}-\eqref{eq:bfm_v}, and its robust counterpart (\ac{LP-RDOE}) by solving \eqref{eq:robust_R+_lp-doe}. The conventional \ac{NLP-DOE} is obtained by solving \eqref{eq:conv_lp-nlp-doe} with the non-linear equations \eqref{eq:pf_bim_ireal}-\eqref{eq:pf_bim_qnode} and its robust counterpart (\ac{NLP-RDOE}) by solving \eqref{eq:robust_D_nlp-doe}.

We have implemented all four approaches from scratch in Julia/JuMP programming language, where GLPK \cite{makhorin2020glpk} is used for solving the \ac{LP} problems and Ipopt \cite{waechter2026ipopt} for solving the \ac{NLP} problems. We validated our implemented power flow models against the established OpenDSS software \cite{dugan2011opendss}. The simulations are run in a HP Elitebook G9 laptop, with a 12th Gen Intel i7-1280P, 1800 Mhz and 32~GB of installed RAM.

\subsection{Conceptual Test Feeder}

The three-phase 2-node conceptual test feeder is based on the widely adopted \ac{MV} \ac{13NF} \cite{kersting2001}. The nominal voltage is 4.16~kV and we select 0.95 and 1.05~p.u. as the voltage limits. The substation voltage is fixed at $\mtx{V}_0=[1 \angle 0^\circ, 1 \angle 240^\circ, 1 \angle 120^\circ]^\top$~p.u. and the three-phase transformer rating is 5~MVA, hence $\overline{S}_{01}^\phi = 1.67$~MVA for each phase \cite{kersting2001}.

The three-phase line between the two nodes is 1.5~km long with configuration 601, emulating the main branch between node 650 and 680 in the \ac{13NF}. We adopt an unbalanced base load with $\widetilde{\mtx{p}}_1=[300, 250, 400]^\top$~kW and $\widetilde{\mtx{q}}_1=[120, 100, 150]^\top$~kvar. This results in the underlying voltage magnitude $|\mtx{V}_1|=[0.986, 0.988, 0.960]^\top$~p.u., noting that phase~c is near the lower limit. For clarity and illustrative purposes, we assume phase~a has no controllable \acp{DER} participating in the \ac{DOE} program (i.e., $p_1^a=0$).

\begin{figure}[!t]
	\centering
	\subfloat[]{%
		\includegraphics[width=0.50\linewidth]{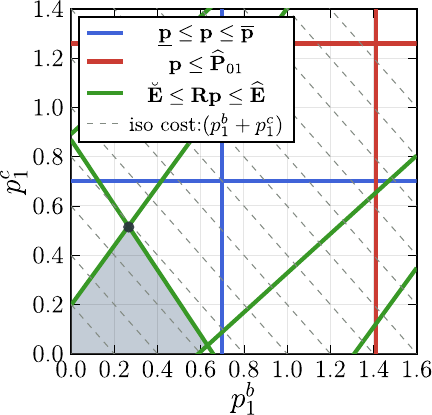}}
	\subfloat[]{%
		\includegraphics[width=0.50\linewidth]{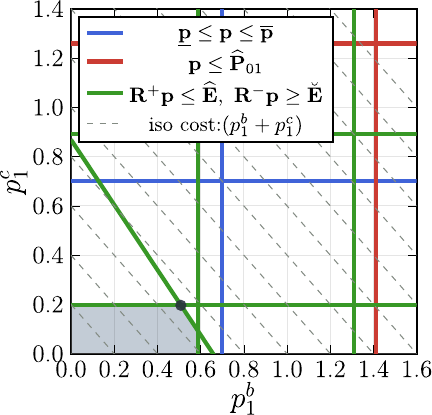}}
    \caption{Feasible region of the import \ac{DOE} optimization with the linear model: (a) conventional (\ac{LP-DOE}); and (b) robust (\ac{LP-RDOE}) calculation.}
	\label{fig:concept_lp-doe_region}
\end{figure}

Fig.~\ref{fig:concept_lp-doe_region} presents the feasible region and the optimal solution (black dot) for the conventional and the robust \ac{LP} approaches. Observe this study case is voltage-constrained as the thermal and the physical envelope constraints are not active in neither case. Upon closing the envelope (box) from the optimal point, part of the envelope will lie outside the feasible region in the \ac{LP-DOE} case, but fully inside in the \ac{LP-RDOE} case.

\begin{figure}[!t]
	\centering
	\includegraphics[width=0.95\linewidth]{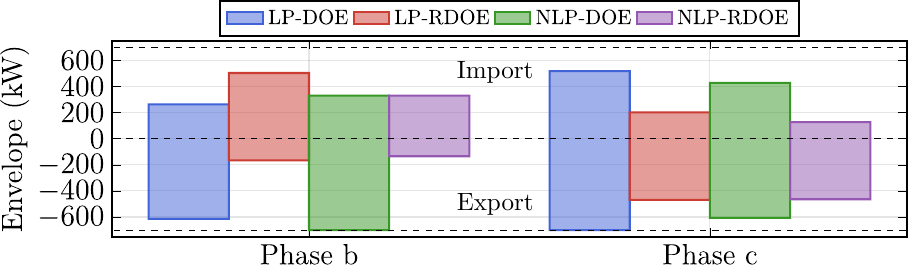}
	\caption{Import and export \ac{DOE} results for the 2-node conceptual feeder.}
	\vspace{-0pt}
	\label{fig:concept_doe_barplot}
\end{figure}

Fig.~\ref{fig:concept_doe_barplot} presents the import and export \acp{DOE} for all four approaches. Observe that both conventional approaches give significantly more import envelope to phase~c than phase~b, although phase~c is much closer to the lower voltage limit. The conventional approaches assume that the negatively-coupled phase~b will always be at 100\% realization, allowing a greater envelope to phase~c. In contrast, the robust approaches give more import envelope to phase~b, as they consider the entire range of realization, including when phase~b realization is zero and no voltage support is provided to phase~c.

Table~\ref{tab:concept_total_doe} presents the total import and export \ac{DOE} for each optimization approach. Observe that the robust approaches provides smaller envelope sizes as they consider the most conservative (worst-case) envelope within the feasible region. The most conservative values are provided by the \ac{NLP-RDOE}, suggesting higher operational security for the grid. Note the difference between \ac{LP-DOE} and \ac{NLP-DOE} is due to the linear approximation of the power flow model.

Table~\ref{tab:concept_doe_results_volt} presents the grid voltage magnitude at node~1 for three realization scenarios by the \acp{AEE}: (i) 100\% realization at both phases b and c; (ii) 50\% at phase~b and 100\% at phase~c; and (iii) 0\% at phase~b and 100\% at phase~c. The numerical results are obtained with the full AC power flow simulation benchmarked against the OpenDSS software. Voltage violations (below 0.95~p.u.) are presented in \textbf{bold}.

In Table~\ref{tab:concept_doe_results_volt}, note that the linear robust approach still results in voltage violations, although in this case the violations stems from the inaccuracy of the linear power flow approximation. The conventional \ac{NLP} results in significant voltage violation at phase~c, as it does not account for uncertain realization. In contrast, the conservative \ac{NLP-RDOE} results in no voltage violations for all simulated scenarios.

\begin{table}[!t]
	\renewcommand{\arraystretch}{1.0}
	\caption{Total import and export \ac{DOE} for the conceptual feeder.}
	\vspace{-10pt}
	\label{tab:concept_total_doe}
	\begin{center}
	\begin{tabular}{|c|c|c|c|c|}
		\hline
		  Case          & LP-DOE    & LP-RDOE     & NLP-DOE     & NLP-RDOE     \\ \hline
		Import (kW)   & 778       & 703         & 755         & 455          \\ \hline
		Export (kW)   & -1314     & -636        & -1307       & -599         \\ \hline
	\end{tabular}
	\end{center}
\end{table}

\begin{table}[!t]
	\renewcommand{\arraystretch}{1.05}
    \setlength{\tabcolsep}{5pt}
	\caption{Voltage magnitude for three import realization scenarios.}
	\vspace{-10pt}
	\label{tab:concept_doe_results_volt}
	\begin{center}
	\begin{tabular}{|c|c|c|c|c|c|}
		\hline
		Realization                                                       & $V$  & LP-DOE         & LP-RDOE        & NLP-DOE        & NLP-RDOE     \\ \hline
		\multirow{3}{*}{\makecell{(i) \\ $u_1^b=p_1^b$ \\ $u_1^c=p_1^c$}}    & a   & 1.012          & 0.974          & 1.001          & 0.977       \\ \cline{2-6} 
								                                       & b   & 0.953          & \textbf{0.943} & 0.950          & 0.960       \\ \cline{2-6} 
								                                       & c   & \textbf{0.936} & 0.982          & 0.950          & 0.975       \\ \hline
		\multirow{3}{*}{\makecell{(ii) \\ $u_1^b=0.5p_1^b$ \\ $u_1^c=p_1^c$}} & a   & 1.020          & 0.987          & 1.011          & 0.986       \\ \cline{2-6} 
								                                       & b   & 0.962          & 0.963          & 0.962          & 0.972       \\ \cline{2-6} 
								                                       & c   & \textbf{0.926} & 0.963          & \textbf{0.937} & 0.962      \\ \hline
        \multirow{3}{*}{\makecell{(iii) \\ $u_1^b=0$ \\ $u_1^c=p_1^c$}}        & a   & 1.029          & 1.002          & 1.021          & 0.996       \\ \cline{2-6} 
								                                       & b   & 0.971          & 0.981          & 0.973          & 0.983       \\ \cline{2-6} 
								                                       & c   & \textbf{0.915} & \textbf{0.944} & \textbf{0.924} & 0.950      \\ \hline
	\end{tabular}
	\end{center}
\end{table}

\subsection{Belgian Residential Distribution Network}

\begin{figure*}[!t]
	\centering
	\includegraphics[width=0.90\linewidth]{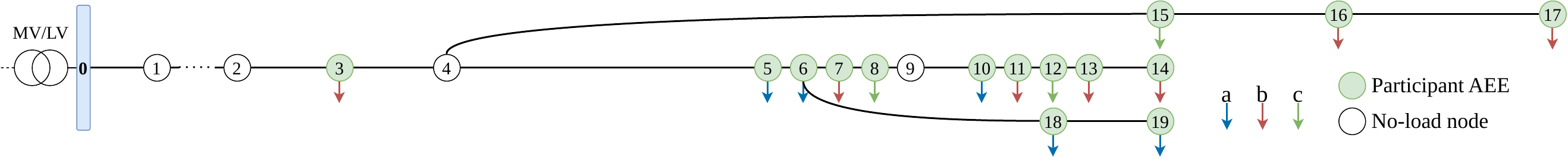}
    \vspace{-5pt}
	\caption{Diagram of the Flobecq three-phase distribution feeder. Horizontal distance from bus~0 represents the electrical distance.}
	\vspace{-5pt}
	\label{fig:flobecq_diagram}
\end{figure*}

Fig.~\ref{fig:flobecq_diagram} illustrates the 20-node \ac{LV} feeder based on a Belgian residential network \cite{klonari2014}. Nominal voltage is $400$~V and the substation voltage is kept balanced at 1~p.u. Voltage bounds are 0.90 and 1.10~p.u. and the substation three-phase transformer is 150~kVA, hence $\overline{S}_{01}^\phi = 50$~kVA for each phase.

Real-world load data from the anonymized Flobecq dataset \cite{klonari2014} is used for the base load $\widetilde{\mtx{p}}$ of the 15 residential \acp{AEE}. A power factor of 0.95 is selected to obtain the base reactive power $\widetilde{\mtx{q}}$. Each \ac{AEE} owns three controllable and participating \acp{DER}: a 5~kWp rooftop solar \ac{PV} system; a 5~kW home battery storage; and a 2.7~kW bidirectional \ac{EV} charger. Those \acp{DER} result in physical envelope bounds $\underline{p}=-12.7$ and $\overline{p}=7.7$ for every \ac{AEE}.

Table~\ref{tab:belg_total_doe} presents the total \ac{DOE} allocated to nodes for a high-demand scenario with limited available grid capacity. Note that export envelope is larger in all cases, reflecting there is more room for generating than consuming additional power. Both robust approaches result in smaller envelopes, \ans{reflecting that grid-safe \acp{DOE} have an impact on the available import/export capacity in which \acp{DSO} need to trade-off.}

%as they take into account different realizations within the envelopes.

Table~\ref{tab:volt_belg_results} shows the number of voltage violations and the lowest voltage magnitude on the network for three distinct realization scenarios. The realization scenarios are: (i) all \acp{AEE} fully utilizing their import \acp{DOE}; (ii) all \acp{AEE} fully utilizing their import \acp{DOE}, except phase~c \acp{AEE} at 0\% realization; and (iii) random import realization for all \acp{AEE}. Observe that both the conventional \ac{LP} and \ac{NLP} led to numerous significant voltage violations in all scenarios, except for the 100\% realization scenario with the \ac{NLP-DOE}, as the optimal \ac{DOE} is obtained from that specific point. In contrast, the robust \ac{NLP-RDOE} presented no voltage violations and the \ac{LP-RDOE} resulted in fewer and milder voltage violations stemming from inaccuracies of the power flow model.

Table~\ref{tab:solve_time} presents the average computation time of each optimization approach over 5 different runs. Observe the \acp{LP} presented shorter computation time with no significant difference between them, as no additional runs or constraints are added to the robust counterpart. The \ac{NLP-RDOE} presented longer computation time as it needs multiple \ac{NLP} runs to obtain the final conservative values, yet shorter than the 5 or 15-min interval that \ac{DOE} engines are usually allowed to run. \ans{See \cite{decarvalho2026cired} for more details on the scalability of \ac{DOE} engines.}

\begin{table}[!t]
	\renewcommand{\arraystretch}{1.0}
	\caption{Total import and export \ac{DOE} for the Belgian feeder.}
	\vspace{-10pt}
	\label{tab:belg_total_doe}
	\begin{center}
	\begin{tabular}{|c|c|c|c|c|}
		\hline
		  Case          & LP-DOE    & LP-RDOE      & NLP-DOE     & NLP-RDOE     \\ \hline
		Import (kW)   & 57.3      & 42.6         & 54.0        & 31.1          \\ \hline
		Export (kW)   & -107.5    & -79.6        & -111.2      & -56.2         \\ \hline
	\end{tabular}
	\end{center}
\end{table}

\begin{table}[!t]
	\renewcommand{\arraystretch}{1.0}
	\caption{Power flow simulations for different realization scenarios.}
	\vspace{-10pt}
	\label{tab:volt_belg_results}
	\begin{center}
	\begin{tabular}{|c|c|c|c|c|}
		\hline
		  Realization         & LP-DOE              & LP-RDOE             & NLP-DOE             & NLP-RDOE     \\ \hline
		(i) $\mtx{u}=\mtx{p}$       & \textbf{27 / 0.893} & 0 / 0.903	        & 0 / 0.900	          & 0 / 0.914    \\ \hline
		(ii) $\mtx{u}^c=0$     & \textbf{16 / 0.850} & \textbf{10 / 0.896}	& \textbf{12 / 0.858} & 0 / 0.919    \\ \hline
        (iii) Random    & \textbf{12 / 0.874} & 0 / 0.913	        & \textbf{12 / 0.880} & 0 / 0.933    \\ \hline
	\end{tabular}
	\end{center}
\end{table}

\begin{table}[!t]
	\renewcommand{\arraystretch}{1.0}
	\caption{Solve time of optimization problems for the Belgian feeder.}
	\vspace{-10pt}
	\label{tab:solve_time}
	\begin{center}
	\begin{tabular}{|c|c|c|c|c|}
		\hline
		  Case        & LP-DOE    & LP-RDOE     & NLP-DOE     & NLP-RDOE     \\ \hline
		Import (s)  & 0.017	    & 0.021	      & 0.239	    & 1.832        \\ \hline
		Export (s)  & 0.021	    & 0.023	      & 0.259	    & 1.823        \\ \hline
	\end{tabular}
	\end{center}
\end{table}

\section{Conclusion}

This paper presents two robust \ac{DOE} calculation approaches for safer operation of unbalanced three-phase distribution systems. The robust \ac{DOE} is formulated to maintain grid feasibility for all envelope realization, rather then only at the envelope bounds as the conventional \acp{DOE}. The proposed linear robust approach keeps the optimization as a single level \ac{LP} problem with the same size as the conventional \ac{LP-DOE}, presenting simple implementation and efficient computation. Numerical results showed that the linear robust improved network security, with remaining violations due to inaccuracies of the linear power flow model that can be improved by a sequential linearization algorithms. We also propose a practical robust approach for the \ac{NLP-DOE} based on representative edge cases that can easily be implemented and adopted by \acp{DSO}. The \ac{NLP-RDOE} led to more conservatives envelopes, but no network violations. Our results support the importance of accurately capturing unbalanced network aspects and realization uncertainty to securely operate distribution networks.

\section*{Acknowledgment}
This work is supported by FNR Luxembourg in the frame of the project DEFIE (INTER/FNRS/24/19145021/DEFIE).


\bibliographystyle{IEEEtran}
\bibliography{./bibtex/mybibfile}

\end{document}